\documentclass[aps,prc,twocolumn,superscriptaddress,showpacs,floatfix,nofootinbib]{revtex4-1}
\usepackage{amsmath,graphicx}

\begin{document}

\title{Directed flow at midrapidity in $
\sqrt{s_{NN}}=2.76$~TeV Pb+Pb collisions}

\author{Ekaterina Retinskaya}
\affiliation{
CEA, IPhT, Institut de physique th\'eorique de Saclay, F-91191
Gif-sur-Yvette, France} 
\author{Matthew Luzum}
\author{Jean-Yves Ollitrault}
\affiliation{
CNRS, URA2306, IPhT, Institut de physique th\'eorique de Saclay, F-91191
Gif-sur-Yvette, France}
\date{\today}

\begin{abstract}
We analyze published data from the ALICE Collaboration in order to obtain the
first extraction of the recently-proposed rapidity-even directed flow
observable $v_1$.  
An accounting of the correlation
due to the conservation of transverse momentum restores the
factorization seen by ALICE in all other Fourier harmonics and thus
indicates that the remaining correlation gives a reliable measurement
of directed flow.  
We then carry out the first viscous hydrodynamic calculation of
directed flow, and show that it is less sensitive to viscosity than
higher harmonics.  
This allows for a direct extraction of the dipole asymmetry of the initial
state, providing a strict constraint on the non-equilibrium dynamics
of the early-time system. 
A prediction is then made for $v_1$ in Au-Au collisions at RHIC. 
\end{abstract}


\maketitle

\section{Introduction}

Azimuthal correlations between particles emitted in heavy-ion
collisions are a useful observable to probe the behavior of 
these systems~\cite{Voloshin:2008dg}. Specifically, one measures the Fourier
coefficient~\cite{Alver:2010gr}
\begin{equation}
V_{n\Delta}\equiv \langle\cos n\Delta\varphi\rangle,
\end{equation}
where 
$\Delta\varphi$ is the relative
azimuthal angle between a pair 
of particles, and $\langle \cdots\rangle$ denotes an average over
pairs and collisions. 
The long-range part of this correlation (defined by a rapidity gap
between the pair) is mostly generated by collective, anisotropic flow
of the strongly-coupled matter created in the
collision~\cite{Luzum:2010sp}.  

The most studied Fourier component is
$V_{2\Delta}$~\cite{Adcox:2002ms,Adler:2002pu,Aamodt:2010pa}, corresponding to
elliptic flow~\cite{Ollitrault:1992bk}. 
Recently it was realized that event-by-event
fluctuations~\cite{Alver:2010gr}, generate a whole series of 
harmonics. This has triggered detailed analyses of $V_{n\Delta}$ for 
$n= $ 3--6~\cite{Adare:2011tg,ALICE:2011ab,Aamodt:2011by,collaboration:2011hfa,Collaboration:2012wg,Sorensen:2011fb}.  

Neglected in these analyses is the first Fourier harmonic $V_{1\Delta}$.  
The observed $V_{1\Delta}$ is smaller than $V_{2\Delta}$ and
$V_{3\Delta}$~\cite{Luzum:2010sp,Aamodt:2011by}, 
and receives a sizable contribution from global momentum 
conservation~\cite{Borghini:2000cm,Borghini:2002mv}
which makes its interpretation less straightforward. 
Fluctuations are expected to create a dipole asymmetry in the
system~\cite{Teaney:2010vd}, resulting in a specific directed flow
pattern, with high transverse momentum particles flowing in the direction of the
steepest gradient and low $p_T$ particles flowing in the opposite
direction. 
Hints of this directed flow have been extracted from published $V_{1\Delta}$ data at
the Relativistic Heavy-Ion Collider (RHIC) by two of the authors~\cite{Luzum:2010fb}, 
and its magnitude and $p_T$-dependence were shown to be in
agreement with ideal hydrodynamic calculations~\cite{Gardim:2011qn}. 
Note that this quantity is distinct from the directed flow observable that has been obtained in the past from 
measurements employing a rapidity-odd projection~\cite{Selyuzhenkov:2011zj}.  
That rapidity-odd $v_1$ gives a negligible contribution to $V_{1\Delta}$ near mid-rapidity 
and represents different physics~\cite{Bozek:2010bi}.

In this Letter, we show that data on $V_{1\Delta}$ obtained by 
ALICE~\cite{Aamodt:2011by} can be explained by the superposition of
two effects: global momentum conservation and directed flow.  
This allows for the first reliable measurement of directed flow at mid rapidity.
We then carry out the first viscous hydrodynamic calculation of directed
flow, and show that these data can be used to constrain the initial
dipole asymmetry of the system for each centrality, putting
strong constraints on models of initial conditions. 

\section{Directed flow from dihadron correlations}

The standard picture of heavy-ion collisions is that an approximately thermalized
fluid is created, which eventually breaks up into particles. Particles
are emitted independently in each event, with an azimuthal
distribution that fluctuates from event to event. 
This yields a two-particle correlation which factorizes into the
product of two single-particle distributions~\cite{Luzum:2011mm}:
\begin{equation}
\label{corrflow}
V_{n\Delta}(p_T^t,p_T^a)=v_n(p_T^t)v_n(p_T^a),
\end{equation}
where the superscripts $t$ and $a$ refer to trigger and associated
particles that can be taken from different bins in transverse momentum, 
and $v_n(p_T)$ is the anisotropic flow coefficient. 
Note that it is possible for the event-averaged correlation to not factorize
even if independent emission holds in each event~\cite{Luzum:2011mm}, 
and it is also possible for intrinsic (``non-flow'') pair correlations to 
factorize~\cite{Kikola:2011tu}.
However, flow is currently the only known mechanism that produces a factorized
correlation in the range of transverse momentum studied here (the bulk of particles).
This factorization has been tested
in Pb-Pb
collisions at the
Large Hadron Collider (LHC)~\cite{Aamodt:2011by,Collaboration:2012wg}: 
this is done by fitting the left-hand side of Eq.~(\ref{corrflow}),
which is a $N\times N$ symmetric matrix for $N$ bins in $p_T$, with
the right-hand side of Eq.~(\ref{corrflow}), using the $N$ values of
$v_n(p_T)$ as fit parameters.  
The ALICE collaboration has shown that, 
while the data do factorize for $n>1$, this
factorization breaks down for $n=1$~\cite{Aamodt:2011by}. 
This is not surprising since there is expected to be an additional long-range correlation
induced by momentum conservation that only affects the first harmonic~\cite{Borghini:2000cm}. The
constraint that all transverse momenta add up to 0 yields a 
back-to-back correlation between pairs, which increases linearly with
the transverse momenta of both particles. This correlation adds to the correlation from flow: 
\begin{equation}
\label{corrv1}
V_{1\Delta}(p_T^t,p_T^a)=v_1(p_T^t)v_1(p_T^a)-kp_T^t p_T^a.
\end{equation} 
\begin{table}
\begin{tabular}{|r|c|c|c|c|}
\hline
Centrality&$\chi^2$, Eq.(\ref{corrflow})&
$\chi^2$, Eq.(\ref{corrv1})&$k$~[$10^{-5}$GeV$^{-2}$]
&$\langle\sum p_T^2\rangle^{-1}$\\
\hline
0--10\% &        
6&
2.0  &       
$2.5^{+1.1}_{-0.3}$&
$6.2$ \\
10--20\%&        
16&
1.7&  
$4.7^{+1.4}_{-0.4}$&
$8.9$ \\
20--30\%&      
45&
2.2 &  
$10.2^{+2.1}_{-0.5}$&  
$13$\\  
30--40\%&    
75&
2.2& 
$20.6^{+3.2}_{-1.6}$& 
$21$ \\ 
40--50\%&     
126& 
2.4 & 
$41.5^{+4.7}_{-3.0}$&  
$35$\\  
\hline
\end{tabular}
\caption{\label{table} 
From left to right: $\chi^2$ per degree of freedom of the fit 
to the ALICE $V_{1\Delta}$~\cite{Aamodt:2011by} 
(restricted to $p_T<4$~GeV/$c$)
using Eq.~(\ref{corrflow}), and using Eq.~(\ref{corrv1}); 
value of $k$ from the fit;
estimated value of $k$ from momentum conservation
in units of $10^{-5}$(GeV/$c)^{-2}$. 
}
\end{table}
(Note that the
  nonflow correlation also factorizes in  this particular
  case~\cite{Kikola:2011tu}, but the sum does not.)
Table~\ref{table} compares the quality of the fit to $V_{1\Delta}$
using Eq.~(\ref{corrflow}) or (\ref{corrv1}).
Adding one single
fit parameter $k$ tremendously increases the quality of the fit for
all centrality windows. We have checked that the values of the fit
parameters depend little on the $p_T$ window.
However, the quality of the fit decreases as higher $p_T$ particles
are included, as observed for other harmonics~\cite{Aamodt:2011by}.
Nevertheless, we include the entire range of values as a systematic uncertainty in 
Table~\ref{table}, varying the lower $p_T$ cutoff between 0.25--0.75 GeV, and the
upper cutoff between 2.5--15.0 GeV.  Similarly, we use this procedure to estimate
a systematic uncertainty in $v_1$ (see Fig.~\ref{fig:alicev1} below).

Next, we check whether the value of $k$ from the fit is 
compatible with the value expected from momentum conservation. 
Assuming for simplicity that momentum conservation is the only source
of correlation, one obtains~\cite{Borghini:2000cm} $k=\langle\sum
p_T^2\rangle^{-1}$,   
where the sum runs over all particles emitted in one event, and
angular brackets denote an average over events in the centrality
class. 
Since experiments measure only charged particles in a restricted
phase-space window, only a rough estimate of this quantity can be made,
by extrapolating from existing data. 
We have used the preliminary identified particle $p_T$ spectra from 
ALICE at midrapidity~\cite{Preghenella:2011jv}, and extrapolated them
outside the $p_T$ acceptance of the detector using Levy
fits~\cite{Aamodt:2011zj}.  
In order to extrapolate to all rapidities, we have 
assumed for simplicity that $p_T$ spectra are independent of rapidity, 
and we have used the total charged multiplicity estimated by the   
ALICE collaboration~\cite{Collaboration:2011rta}.
Neutral particles were taken into account assuming isospin symmetry,
and the contribution of particles heavier than nucleons was
neglected. 

The resulting estimate is shown in the last column of
Table~\ref{table}. The fit result in general has the correct size and
increases with \% centrality, as expected. The centrality dependence
is steeper  
than expected from our rough estimate, however --- the fit value 
is larger than the estimated value for the most peripheral bin, while it is significantly smaller
for central collisions.  We cannot explain this, but overall the agreement is 
reasonable, and a discrepancy in $k$ of this size has a very small effect on the extracted directed flow;
the extracted directed flow curves with $k$ fixed to the estimated values were also included in the systematic
error band in $v_1$, but only have a small effect on the two most central bins.

It has been suggested that the correlation from momentum conservation
could be larger than our estimate 
because of approximate conservation of transverse momentum
within smaller subsystems of the entire collision system---specifically rapidity slices of roughly unit
extent~\cite{Borghini:2006yk,Chajecki:2008yi}. 
However, we see no evidence here for such an enhancement.

Thus, by taking into account the only obvious non-flow correlation, the factorization seen 
in higher harmonics is restored, and we can take the resulting $v_1(p_T)$ as a reliable 
measurement of directed flow $v_1$ (presented in Fig.~\ref{fig:alicev1} below).

\section{Results of Hydrodynamic Calculations}

Relativistic viscous hydrodynamics has been shown to successfully
reproduce $v_n$ for $n=2,3,4$~\cite{Schenke:2011bn}. 
Here, we present the first viscous hydrodynamic calculation for
directed flow, $v_1$. 
In hydrodynamics, $v_1$ and the corresponding event-plane angle
$\Psi_1$ are defined by $v_1e^{i\Psi_1}\equiv\langle 
e^{i\varphi}\rangle$, where angular brackets denote an average over
the momentum distribution at freeze-out~\cite{Kolb:2003dz}. 
A collision of identical nuclei at mid-rapidity has $\varphi\to\varphi+\pi$ 
symmetry except for fluctuations, hence $v_1$ at midrapidity is 
solely due to event-by-event fluctuations in the initial state. 

In event-by-event ideal hydrodynamic calculations, $v_1$ was found~\cite{Gardim:2011qn} to be 
approximately proportional to the dipole asymmetry of the system $\varepsilon_1$
defined as~\cite{Teaney:2010vd}
\begin{equation}
\varepsilon_1\equiv \frac{\left|\{r^3 e^{i\phi}\}\right|}{\{r^3\}}.
\end{equation}
where $\{\cdots\}$ denotes an average value over the initial energy density after recentering the coordinate system ($\{r
e^{i\phi}\}=0$).  

\begin{figure}
 \includegraphics[width=\linewidth]{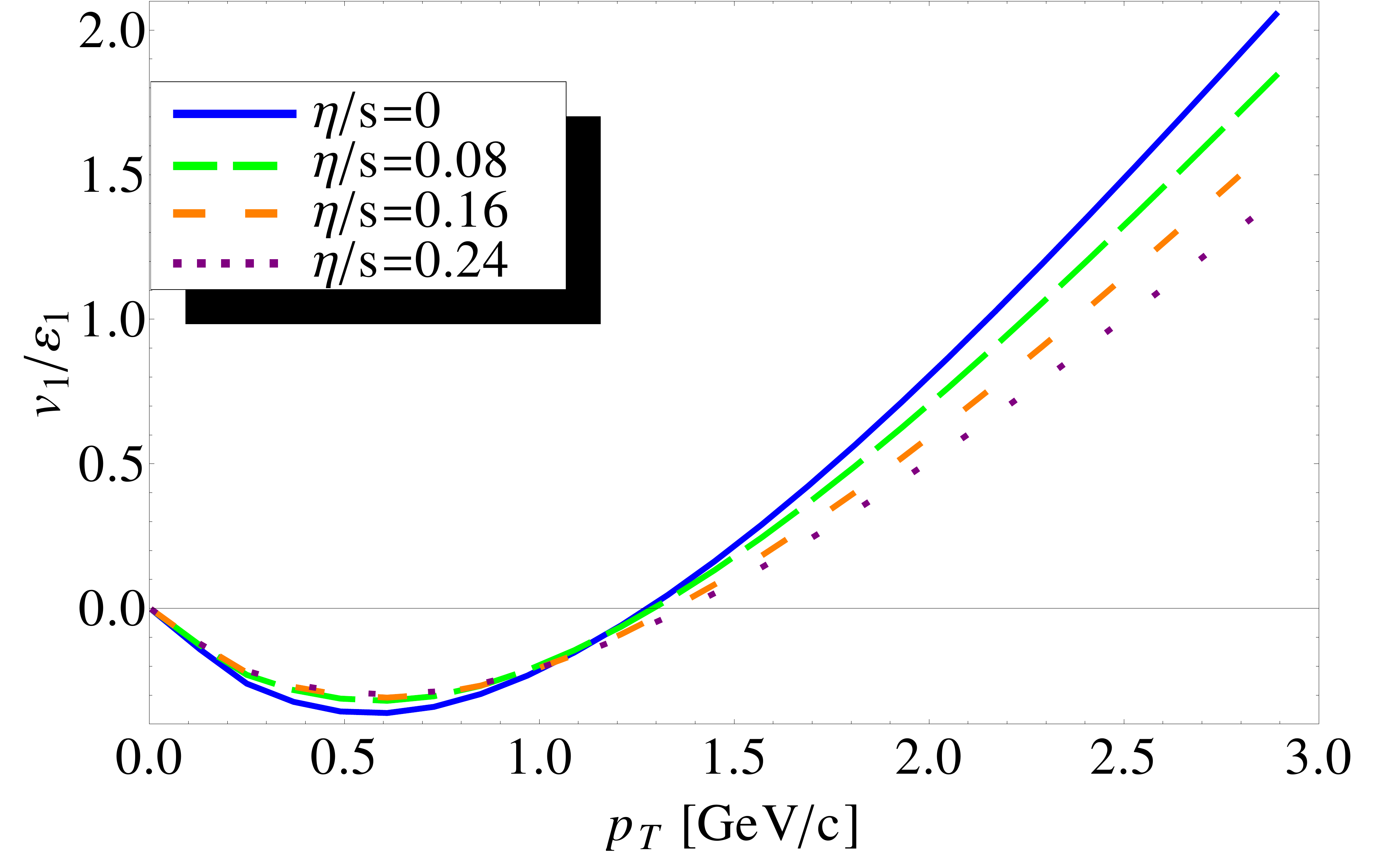}
 \caption{
(Color online) 
Directed flow $v_1$, scaled by the initial dipole asymmetry
$\varepsilon_1$, in a central Pb-Pb
collision at 2.76~TeV, for different values of the shear viscosity to
entropy ratio $\eta/s$.  
}  
\label{fig:v1hydro}
\end{figure}
Here, in order to make a systematic study, we use a smooth, symmetric density profile which we 
deform to introduce a dipole asymmetry of the desired size and orientation. 
Specifically, our calculation is a 2+1 dimensional viscous
hydrodynamic calculation which uses as initial condition the transverse 
energy density ($\epsilon(r,\phi)$) profile from an optical Glauber
model~\cite{Luzum:2009sb}, which is deformed in a way analogous to the previous study of $v_3$ and higher harmonics in Ref.~\cite{Alver:2010dn}: 
\begin{equation}
\epsilon(r,\phi)\rightarrow\epsilon\left(r\sqrt{1+\delta\cos(\phi-\Phi_1)},\phi\right),
\end{equation}
where $\delta$ is a small parameter. 
Both $v_1$ and $\varepsilon_1$ are proportional to $\delta$ 
for $\delta\ll 1$. 
For noncentral collisions, $v_1$ depends mildly on the orientation of 
the dipole asymmetry $\Phi_1$ with respect to the impact parameter.
Our results are averaged over $\Phi_1$.  

Fig.~\ref{fig:v1hydro} presents the ratio $v_1/\varepsilon_1$ as a
function of the transverse momentum $p_T$ for central collisions. 
Unlike higher-order harmonics, which are usually positive for all
$p_T$, $v_1$ changes sign. The reason is that the net transverse
momentum of the system is zero by construction, which implies $\langle
p_T v_1(p_T)\rangle=0$: low-$p_T$ particles tend to flow in the direction
opposite to high-$p_T$ particles. 

The harmonics $v_n$ tend to probe smaller length scales with increasing $n$,
and as a result are expected to have an increasing sensitivity to viscosity.
Our results show that, indeed, $v_1$ is less sensitive to viscosity than
$v_2$~\cite{Luzum:2008cw} and higher
harmonics~\cite{Alver:2010dn,Schenke:2011bn}.
This insensitivity to viscosity combined with the approximate proportionality 
$v_1\propto\varepsilon_1$ provides a unique opportunity to place a direct constraint
on the dipole asymmetry of the early-time system.

In a realistic Pb-Pb collision, $\varepsilon_1$ varies from event to
event. 
The contribution of directed flow to $V_{1\Delta}$ scales like
$\varepsilon_1^2$. 
Therefore the experimentally measured $v_1$ scales like the
root-mean-square (rms) value of $\varepsilon_1$ in the centrality
bin. As we shall see below, there is a wide range of predictions for
this quantity.  With these new data we can now quickly discern which are compatible with experiment
by identifying an allowed range of values for the rms dipole asymmetry.

\begin{figure}
\includegraphics[width=\linewidth]{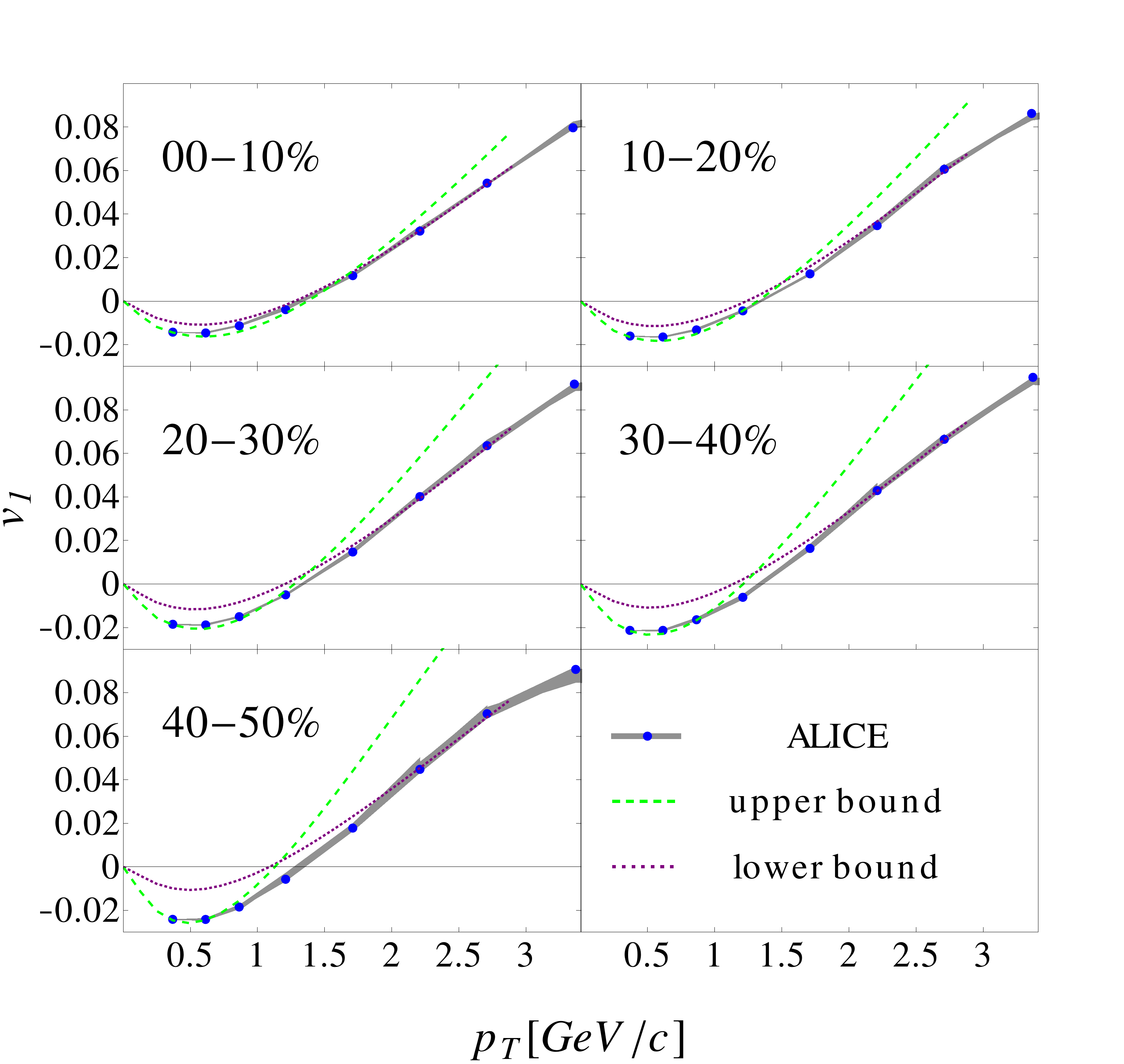}
 \caption{
(Color online) $v_1(p_T)$ in Pb-Pb collisions at 2.76~TeV extracted
from correlation data~\cite{Aamodt:2011by}, in various centrality
windows.  The shaded band represents the systematic uncertainty from
the choice of $p_T$ window used for the fit. 
The curves are hydrodynamic calculations where the value of
$\varepsilon_1$ has been adjusted so as to match the data from above
or below, and which were used to obtain the upper and lower bound, respectively, in Fig.~\ref{fig:epsilon}. }  
\label{fig:alicev1}
\end{figure}
Fig.~\ref{fig:alicev1} displays $v_1$ versus $p_T$ extracted from
ALICE correlation data using Eq.~(\ref{corrv1}). 
The magnitude and $p_T$ dependence of $v_1$ are similar at LHC and at
RHIC~\cite{Luzum:2010fb}, and the mild centrality dependence, reminiscent 
of $v_3$~\cite{Adare:2011tg,ALICE:2011ab}, is expected since both are generated
purely from fluctuations in the initial state.

The $p_T$ dependence of $v_1$ in LHC data bears a striking resemblance to that predicted by
hydrodynamics, Fig.~\ref{fig:v1hydro}. 
In a given centrality window and for a given value of the viscosity,
one can tune the value of the  dipole asymmetry $\varepsilon_1$ in the
hydrodynamic calculation so as  to obtain reasonable agreement with
data.  
If one chooses to match data at the lowest $p_T$, calculation overpredicts
data at high $p_T$. Conversely, if one matches data at high $p_T$,
calculation underpredicts data at low $p_T$. 
The corresponding values of $\varepsilon_1$ can be considered upper
and lower bounds on the actual value. 

The values of $\eta/s$ (the ratio of shear viscosity to entropy density) implied by
comparisons of elliptic flow data to hydrodynamic calculations all lie in the
range $0<\eta/s<0.24$~\cite{Luzum:2009sb,Schenke:2011tv,Shen:2011eg}.
Assuming that $\eta/s$ lies in this range, we can extract an allowed range
for the dipole asymmetry, using the extremal values of $\eta/s$ and $\varepsilon_1$
that still give a reasonable fit to data.
 (the extremal curves used are shown in Fig.~\ref{fig:alicev1}).
\begin{figure}
 \includegraphics[width=\linewidth]{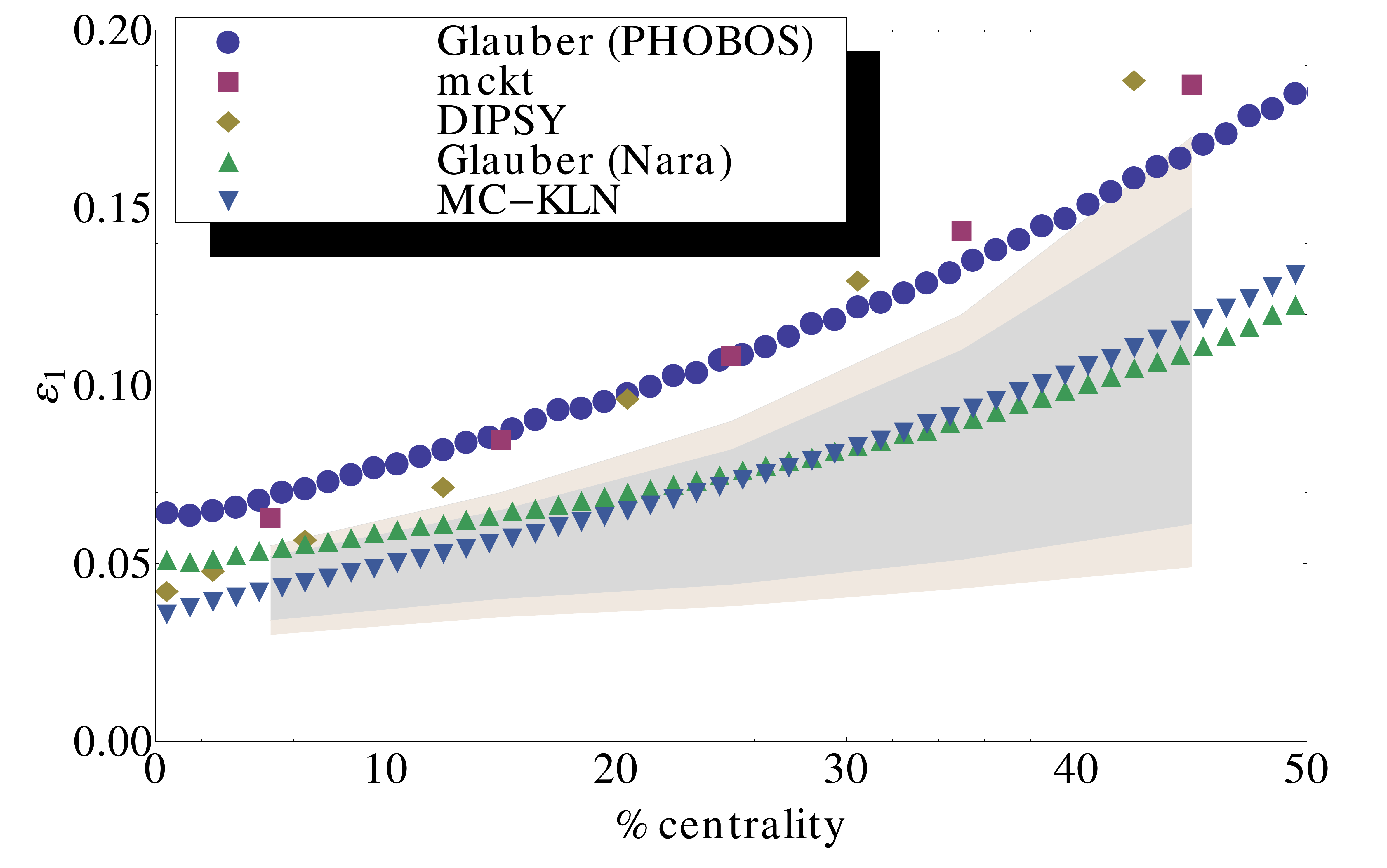}
 \caption{
(Color online) Variation of $\varepsilon_1$ with centrality.
The shaded bands indicate the allowed regions using ALICE data in
combination with viscous hydrodynamics, assuming either
$0<\eta/s<0.24$ (lighter shade) or 
$0.08<\eta/s<0.16$ (darker shade). 
Also shown are the prediction of several Monte-Carlo models of initial
conditions: DIPSY~\cite{Flensburg:2011wx},
MC-KLN (obtained from mckt v1.00)~\cite{Albacete:2010ad, mckt},
mckt v1.25~\cite{Dumitru:2012yr, mckt}, 
and two different implementations of the Glauber
model utilizing point-like nucleons~\cite{Alver:2008aq} or uniform disks \cite{mckt}.  The binary collision fraction for both Glauber models was taken to be 
$x=0.18$~\cite{Bozek:2010wt}. 
}  
\label{fig:epsilon}
\end{figure}

Fig.~\ref{fig:epsilon} displays the allowed
values of $\varepsilon_1$ as a function of centrality, together with
the rms $\varepsilon_1$ from various Monte-Carlo models of initial
conditions.  The allowed range assuming $0.08<\eta/s<0.16$, representing
the most common values extracted from $v_2$ data, are also shown in a darker band,
to illustrate the small effect of viscosity.
Both the order of magnitude and the centrality dependence of
$\varepsilon_1$ from Monte-Carlo models resemble the allowed values
from LHC data. 
However, there are significant differences between the models. 
LHC data already exclude DIPSY~\cite{Flensburg:2011wx} above 10\%
centrality, 
and the Phobos Glauber model~\cite{Alver:2008aq}  as well as a recent
improved mckt model with KNO fluctuations~\cite{Dumitru:2012yr} 
over the entire centrality range. 

Note, however, that since hydrodynamics is expected to be more reliable at low 
transverse momentum, the actual value of $\varepsilon_1$ is most likely to lie very close 
to our upper bound.  Thus it is possible that the two models with the largest 
dipole asymmetry may only need a slight tuning to achieve a correct value, while
the value from the others may in fact be too low.  We prefer here to be conservative in our 
claimed region of allowed values and leave it to future study for more stringent conclusions.

\begin{figure}
 \includegraphics[width=\linewidth]{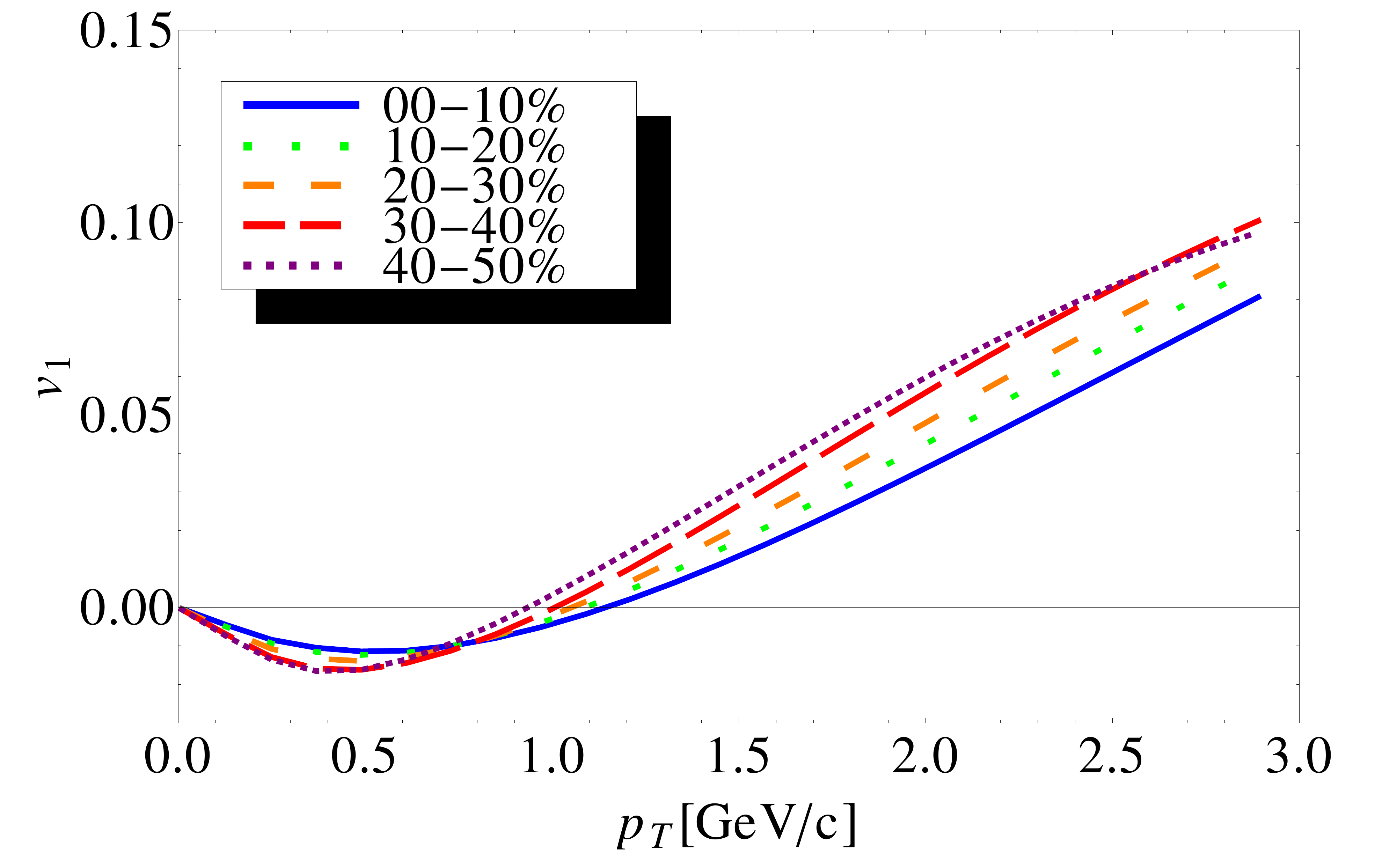}
 \caption{
(Color online) Viscous hydrodynamic prediction for $v_1$ in Au-Au
collisions at 200 GeV in various centrality windows.}  
\label{fig:rhic}
\end{figure}
Models such as those presented here do not in general predict a significant
change in dipole asymmetry with collision energy.  By extracting a best value 
of $\varepsilon_1$ from these LHC data combined with hydrodynamic calculations
of lower energy collisions, we can make predictions for Au-Au collisions at RHIC 
assuming little change in the average dipole asymmetry in a centrality bin.
Since the change of $\varepsilon_1$ with collision energy predicted by each current
Monte Carlo model is much smaller than the range spanned by the various models,
this prediction is more reliable than any obtained by assuming a particular model for the initial conditions.
These are presented in Fig.~\ref{fig:rhic}.
The value of $\varepsilon_1$ at LHC is obtained
by taking the best fit to the experimental $v_1$ for
$p_T<1.5$~GeV/c, which is the range where hydrodynamics agrees best
with data~\cite{Kolb:2003dz}. 
Our calculations use $\eta/s=0.16$ both at LHC and at RHIC, but 
the extrapolation from LHC to RHIC depends very weakly on the assumed value
of $\eta/s$~\cite{Luzum:2009sb}.  

These predictions are compatible with the
attempted extraction of $v_1$~\cite{Luzum:2010fb} from a much more limited 
set of correlation data at 20--60\% centrality released by the STAR collaboration,
but a dedicated analysis by one of the experimental collaborations at RHIC
will allow for a much more precise test, including centrality dependence.

\section{Conclusions}

We have shown that the first Fourier component of the two-particle
azimuthal correlation measured at LHC, $V_{1\Delta}$, can be explained by
collective flow, much in the same way as higher harmonics, after the
correlation from momentum conservation is accounted for. 
We have thus obtained the first measurement of directed flow, $v_1$,
at midrapidity at the LHC. 
This experimental result was compared with the first viscous
hydrodynamic calculation of directed flow. $v_1$ 
was found to have a weaker dependence on viscosity than $v_2$ and $v_3$,  
which allows for the first time a tight constraint to be placed directly on the geometry
and fluctuations of the early-time system, and which rules out certain current theoretical models.
The extracted values of the dipole asymmetry of the initial conditions then allow for 
predictions to be made for directed flow at midrapidity in 
lower-energy collisions at RHIC, which were presented.

\begin{acknowledgments}
For providing experimental data from the ALICE Collaboration we would like to thank Andrew Adare and Mateusz Ploskon.
 ML is supported by the European Research Council under the
Advanced Investigator Grant ERC-AD-267258.
\end{acknowledgments}


\begin{thebibliography}{99}

\bibitem{Voloshin:2008dg}
  S.~A.~Voloshin, A.~M.~Poskanzer and R.~Snellings,
  arXiv:0809.2949 [nucl-ex].

\bibitem{Alver:2010gr}
  B.~Alver and G.~Roland,
  Phys.\ Rev.\  C {\bf 81}, 054905 (2010)
  [Erratum-ibid.\  C {\bf 82}, 039903 (2010)]
  [arXiv:1003.0194 [nucl-th]].

\bibitem{Luzum:2010sp}
  M.~Luzum,
  Phys.\ Lett.\  B {\bf 696}, 499 (2011)
  [arXiv:1011.5773 [nucl-th]].

\bibitem{Adcox:2002ms}
  K.~Adcox {\it et al.}  [PHENIX Collaboration],
  Phys.\ Rev.\ Lett.\  {\bf 89}, 212301 (2002)
  [arXiv:nucl-ex/0204005].

\bibitem{Adler:2002pu}
  C.~Adler {\it et al.}  [STAR Collaboration],
  Phys.\ Rev.\  C {\bf 66}, 034904 (2002)
  [arXiv:nucl-ex/0206001].

\bibitem{Aamodt:2010pa}
  K.~Aamodt {\it et al.}  [The ALICE Collaboration],
  Phys.\ Rev.\ Lett.\  {\bf 105}, 252302 (2010)
  [arXiv:1011.3914 [nucl-ex]].

\bibitem{Ollitrault:1992bk}
  J.~Y.~Ollitrault,
  Phys.\ Rev.\  D {\bf 46}, 229 (1992).

\bibitem{Adare:2011tg}
  A.~Adare {\it et al.}  [PHENIX Collaboration],
  Phys.\ Rev.\ Lett.\  {\bf 107}, 252301 (2011)
  [arXiv:1105.3928 [nucl-ex]].

\bibitem{ALICE:2011ab}
  C.~A.~Loizides {\it et al.}  [ALICE Collaboration],
  Phys.\ Rev.\ Lett.\  {\bf 107}, 032301 (2011)
  [arXiv:1105.3865 [nucl-ex]].

\bibitem{Aamodt:2011by}
  K.~Aamodt {\it et al.}  [ALICE Collaboration],
  Phys.\ Lett.\  B {\bf 708}, 249 (2012)
  [arXiv:1109.2501 [nucl-ex]].

\bibitem{collaboration:2011hfa}
  J.~Jia [ATLAS Collaboration],
  J.\ Phys.\ G {\bf 38}, 124012 (2011)
  [arXiv:1107.1468 [nucl-ex]].

\bibitem{Collaboration:2012wg} 
 S.~Chatrchyan {\it et al.}  [CMS Collaboration],
  Eur.\ Phys.\ J.\ C {\bf 72}, 2012 (2012)
  [arXiv:1201.3158 [nucl-ex]].

\bibitem{Sorensen:2011fb}
  P.~Sorensen  [STAR Collaboration],
  J.\ Phys.\ G {\bf 38}, 124029 (2011)
  [arXiv:1110.0737 [nucl-ex]].

\bibitem{Borghini:2000cm}
  N.~Borghini, P.~M.~Dinh and J.~Y.~Ollitrault,
  Phys.\ Rev.\  C {\bf 62}, 034902 (2000)
  [arXiv:nucl-th/0004026].

\bibitem{Borghini:2002mv}
  N.~Borghini, P.~M.~Dinh, J.~Y.~Ollitrault, A.~M.~Poskanzer and S.~A.~Voloshin,
  Phys.\ Rev.\  C {\bf 66}, 014901 (2002)
  [arXiv:nucl-th/0202013].

\bibitem{Teaney:2010vd}
  D.~Teaney and L.~Yan,
  Phys.\ Rev.\  C {\bf 83}, 064904 (2011)
  [arXiv:1010.1876 [nucl-th]].

\bibitem{Luzum:2010fb}
  M.~Luzum and J.~Y.~Ollitrault,
  Phys.\ Rev.\ Lett.\  {\bf 106}, 102301 (2011)
  [arXiv:1011.6361 [nucl-ex]].

\bibitem{Gardim:2011qn}
  F.~G.~Gardim, F.~Grassi, Y.~Hama, M.~Luzum and J.~Y.~Ollitrault,
  Phys.\ Rev.\  C {\bf 83}, 064901 (2011)
  [arXiv:1103.4605 [nucl-th]].

\bibitem{Selyuzhenkov:2011zj} 
  I.~Selyuzhenkov,
  J.\ Phys.\ G G {\bf 38}, 124167 (2011)
  [arXiv:1106.5425 [nucl-ex]].

\bibitem{Bozek:2010bi} 
  P.~Bozek and I.~Wyskiel,
  Phys.\ Rev.\ C {\bf 81}, 054902 (2010)
  [arXiv:1002.4999 [nucl-th]].

\bibitem{Luzum:2011mm}
  M.~Luzum,
  J.\ Phys.\ G {\bf 38}, 124026 (2011)
  [arXiv:1107.0592 [nucl-th]].

\bibitem{Kikola:2011tu}
  D.~Kikola, L.~Yi, S.~Esumi, F.~Wang and W.~Xie,
  arXiv:1110.4809 [nucl-ex].

\bibitem{Preghenella:2011jv}
  R.~Preghenella  [for the ALICE Collaboration],
  Acta Phys.\ Polon.\ B {\bf 43}, 555 (2012)
  [arXiv:1111.7080 [hep-ex]].

\bibitem{Aamodt:2011zj}
  K.~Aamodt {\it et al.}  [ALICE Collaboration],
  Eur.\ Phys.\ J.\  C {\bf 71}, 1655 (2011)
  [arXiv:1101.4110 [hep-ex]].

\bibitem{Collaboration:2011rta} 
  A.~Toia,
  J.\ Phys.\ G G {\bf 38}, 124007 (2011)
  [arXiv:1107.1973 [nucl-ex]].

\bibitem{Borghini:2006yk} 
  N.~Borghini,
  Phys.\ Rev.\ C {\bf 75}, 021904 (2007)
  [nucl-th/0612093].

\bibitem{Chajecki:2008yi} 
  Z.~Chajecki and M.~Lisa,
  Phys.\ Rev.\ C {\bf 79}, 034908 (2009)
  [arXiv:0807.3569 [nucl-th]].


\bibitem{Schenke:2011bn} 
  B.~Schenke, S.~Jeon and C.~Gale,
  Phys.\ Rev.\ C {\bf 85}, 024901 (2012)
  [arXiv:1109.6289 [hep-ph]].

\bibitem{Kolb:2003dz}
  P.~F.~Kolb and U.~W.~Heinz,
  arXiv:nucl-th/0305084.

\bibitem{Luzum:2009sb}
  M.~Luzum and P.~Romatschke,
  Phys.\ Rev.\ Lett.\  {\bf 103}, 262302 (2009)
  [arXiv:0901.4588 [nucl-th]].

\bibitem{Alver:2010dn}
  B.~H.~Alver, C.~Gombeaud, M.~Luzum and J.~Y.~Ollitrault,
  Phys.\ Rev.\  C {\bf 82}, 034913 (2010)
  [arXiv:1007.5469 [nucl-th]].

\bibitem{Luzum:2008cw}
  M.~Luzum and P.~Romatschke,
  Phys.\ Rev.\  C {\bf 78}, 034915 (2008)
  [Erratum-ibid.\  C {\bf 79}, 039903 (2009)]
  [arXiv:0804.4015 [nucl-th]].

\bibitem{Schenke:2011tv}
  B.~Schenke, S.~Jeon and C.~Gale,
  Phys.\ Lett.\  B {\bf 702}, 59 (2011)
  [arXiv:1102.0575 [hep-ph]].

\bibitem{Shen:2011eg}
  C.~Shen, U.~Heinz, P.~Huovinen and H.~Song,
  Phys.\ Rev.\  C {\bf 84}, 044903 (2011)
  [arXiv:1105.3226 [nucl-th]].


\bibitem{Flensburg:2011wx}
  C.~Flensburg,
  arXiv:1108.4862 [nucl-th].

\bibitem{Albacete:2010ad}
  J.~L.~Albacete and A.~Dumitru,
  arXiv:1011.5161 [hep-ph].

\bibitem{mckt}
Code by A.~Dumitru, a fork of MC-KLN by Y.~Nara.  Versions 1.00 and 1.25 obtained from \url{http://physics.baruch.cuny.edu/files/CGC/CGC_IC.html}

\bibitem{Dumitru:2012yr} 
  A.~Dumitru and Y.~Nara,
  Phys.\ Rev.\ C {\bf 85}, 034907 (2012)
  [arXiv:1201.6382 [nucl-th]].

\bibitem{Alver:2008aq}
  B.~Alver, M.~Baker, C.~Loizides and P.~Steinberg,
  arXiv:0805.4411 [nucl-ex].


\bibitem{Bozek:2010wt}
  P.~Bozek, M.~Chojnacki, W.~Florkowski, B.~Tomasik,
  Phys.\ Lett.\  {\bf B694}, 238-241 (2010).
  [arXiv:1007.2294 [nucl-th]].


\end{thebibliography}
\end{document}